\documentclass[
    final            
  ]
  {aipproc}

\layoutstyle{6x9}

\begin{document}

\title{Self-Organizing Maps. An application to the OGLE data and the Gaia Science Alerts}


\author{{\L}ukasz Wyrzykowski}{
  address={Institute of Astronomy, University of Cambridge, UK},
  altaddress={Warsaw University Astronomical Observatory, Poland}
}

\author{Vasily Belokurov}{
  address={Institute of Astronomy, University of Cambridge, UK}
}

\begin{abstract}

Self-Organizing Map (SOM) is a promising tool for exploring large
multi-dimensional data sets. It is quick and convenient to train in an
unsupervised fashion and, as an outcome, it produces natural clusters
of data patterns.  An example of application of SOM to the new
OGLE-III data set is presented along with some preliminary results.

Once tested on OGLE data, the SOM technique will also be implemented
within the Gaia mission's photometry and spectrometry analysis, in
particular, in so-called classification-based Science Alerts.  SOM
will be used as a basis of this system as the changes in brightness
and spectral behaviour of a star can be easily and quickly traced on a
map trained in advance with simulated and/or real data from other
surveys.
\end{abstract}

\maketitle


\section{SOM - what is it?}
The Self-Organizing Map is described by its author Teuvo Kohonen
(\cite{Kohonen}) as a map reflecting topological ordering. It is a
list of weight vectors organised as a 2D grid of map nodes
(neurones). Each datum is mapped onto a node associated with the
nearest weight vector, e.g. the one with the smallest Euclidean
distance from the data pattern, but any kind of similarity measure can
be used.  The SOM organises itself during a competitive and
unsupervised learning process.  Each pattern is shown to the SOM
(randomly or sequentially) and the closest node ("winner") is
found. Then all the neighbouring nodes of the winner are adjusted with
the learning rate $\alpha$ :
 
\begin{equation}
\mathbf{m}_i(t+1) = \mathbf{m}_i(t) + \alpha ( \mathbf{x}(t) - \mathbf{m}_i(t) )
\end{equation}

In the next step, the neighbourhood radius and the learning rate is
decreased and next pattern is shown. The process continues until the
set of patterns is exhausted or the learning rate reaches 0.

\subsection{Describing a pattern}

This is the crucial and the most challenging step when working with
SOMs. The pattern should be described in the most efficient way,
however, with emphasis on characteristic features of the classes. The
advantage of SOMs is that patterns can be as long as we like. For
example, a pattern can contain a whole picture, a binned light curve
or a spectrum, a vector of statistical parameters, binned periodogram,
or the combination of these.

\begin{figure}
  \includegraphics[height=.17\textheight]{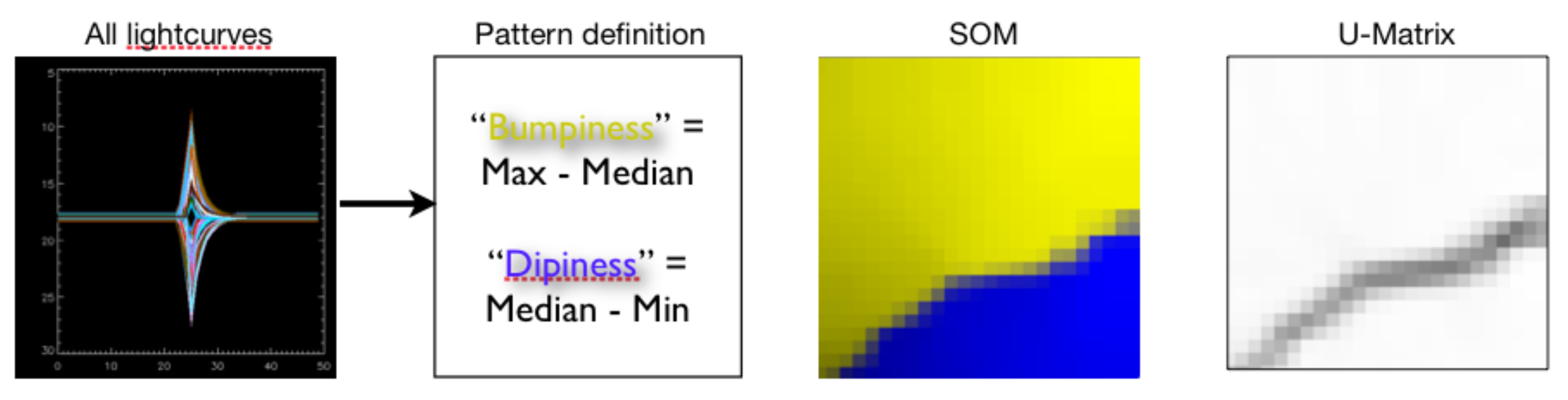}
  \caption{The simple example of SOM. Two kinds of light curves - a
    bump and a dip with different amplitudes - are described by two
    parameters, therefore each pattern contains two values:
    "bumpiness" and "dipiness". Resulting SOM is visualised by
    assigning its vectors to the RGB channels (here Red and Green
    channels were joined). U-Matrix plot shows the distribution of
    distances between the nodes of the SOM. This SOM easily
    disentangles the two light curve classes.}
\end{figure}

\begin{figure}[!b]
  \includegraphics[height=.40\textheight]{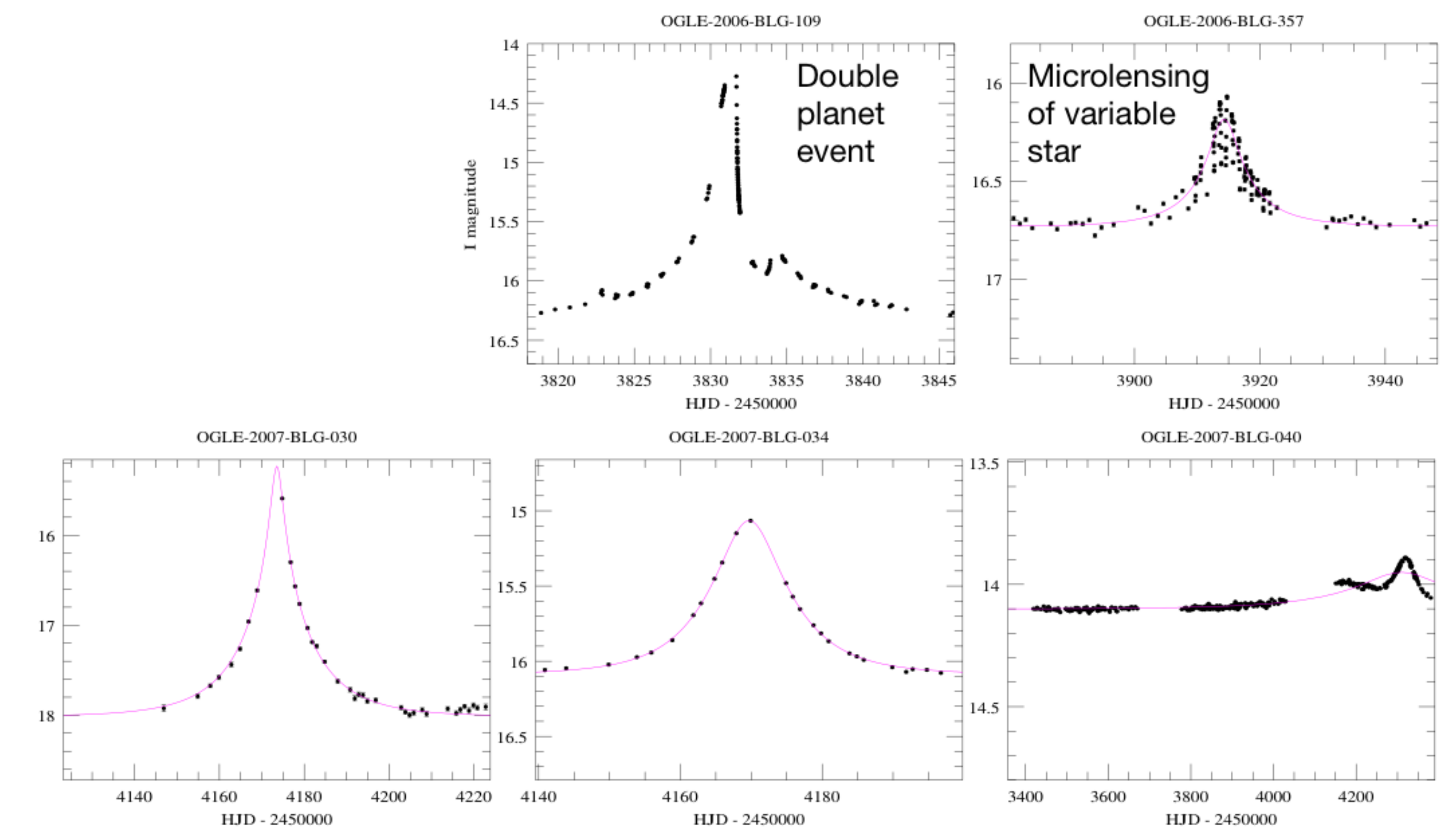}
  \caption{Examples of OGLE-III microlensing events with standard model curve (magenta line).}
\label{fig:2}
\end{figure}

\section{Optical Gravitational Lensing Experiment}
OGLE has run since 1992 at the University of Warsaw (Poland). It uses
a dedicated 1.3m telescope in Chile, which continuously monitors
hundreds of millions of stars towards the Galactic Centre and
Magellanic Clouds in order to detect gravitational microlensing events
(\cite{Paczynski1996}).  As a natural by-product of such search, vast
number of variable stars are being discovered and monitored.
Presently, nearly a billion objects towards the Galactic bulge and
Magellanic Clouds need to be investigated and classified into
variability classes\footnote{For the first attempt of applying SOM to the variable
stars detected by OGLE see http://www.ast.cam.ac.uk/$\sim$vasily/ogle\_som/}.

In early 2009 OGLE will upgrade its camera to 34 CCDs covering about 1
sq. deg. in one exposure. It will create vast data sets which will
have to be analysed in an automated manner.

OGLE's Early Warning System (EWS) \cite{Udalski2003} is detecting
on-going microlensing event almost in real-time. Since 2002 it has
detected nearly 4000 candidates for events. Most of them are following
typical Paczynski's curve (see Fig. \ref{fig:2}), but there are also
anomalous events due to e.g. binary lens, parallax effect or presence
of a planet (e.g. \cite{Gaudi2008}). These anomalous light curves vary
enormously in their shape and present a challenge for a quick and
robust classification.

\begin{figure}[!b]
  \includegraphics[height=.45\textheight]{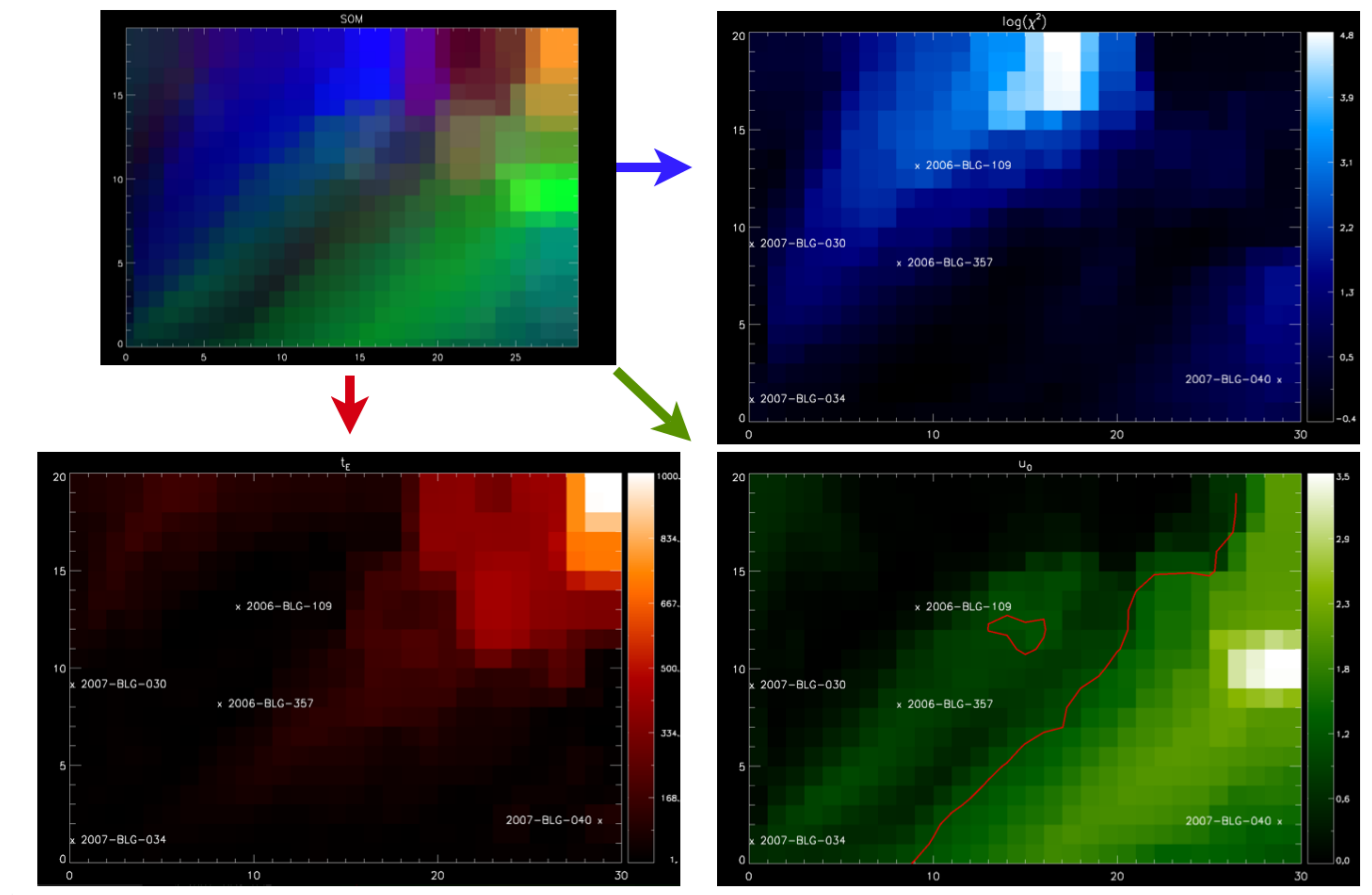}
  \caption{SOM trained on OGLE-III microlensing events parameters ($t_E$, $u_0$ and $\chi ^2$). The SOM (upper left) can be also visualised by spliting each RGB channel separately for each parameter. Positions of exemplary microlensing events (see \ref{fig:2}) are marked. Boundary of events with lens crossing the Einstein Radius ($u_0$<1) is marked as a red contour.   }
  \label{fig:3}
\end{figure}

The SOM shown in Figure \ref{fig:3} was trained on patterns containing
three microlensing model parameters: time-scale ($t_E$), impact
parameter ($u_0$) and goodness of the model fit ($log \chi^2$). Every
parameter was coded in different RBG channel and three maps show each
channel separately.  Positions of exemplary events are marked on the
maps. Anomalous and spurious events have high $\chi^2$
(e.g. 2006-BLG-109). Such trained
map can be used for a quick visualisation of characteristics of new
events detected by EWS.

\section{Gaia}
Gaia is the European Space Agency's corner-stone mission, aiming
mainly at all-sky high-precision astrometry of stars down to V=20
mag. It will also collect spectrometry and photometry data of billions
of stars.  Over its five years of operation, Gaia will scan the entire
sky and will return to each place, on average, around 80
times. However, some places of the sky (around the nodes of the Gaia's
spinning axes) will be observed up to 250 times.

\begin{figure}[!b]
  \includegraphics[height=.55\textheight]{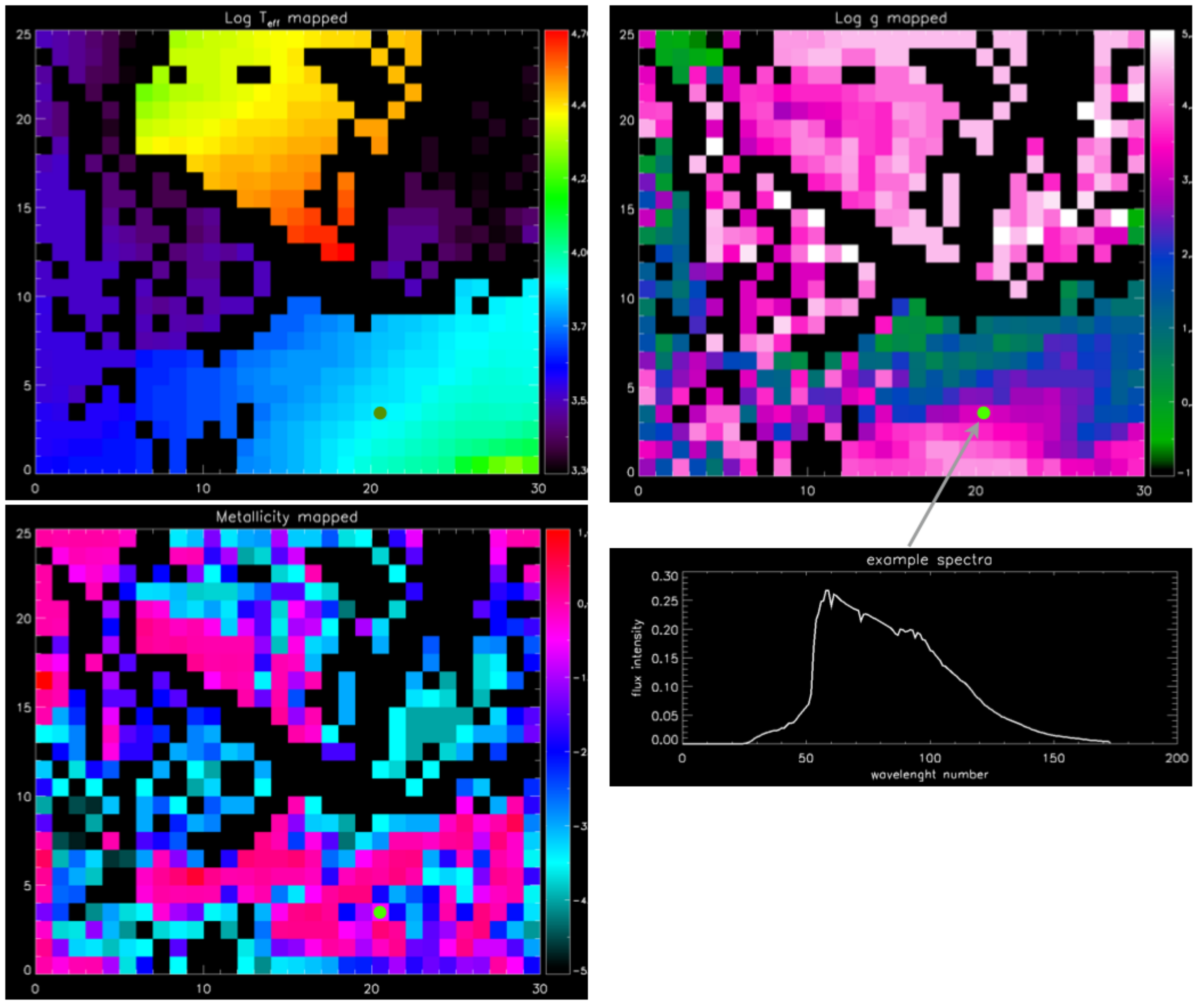}
  \caption{Maps of Basel library spectra physical parameters mapped back on the SOM. Any spectrum shown to the SOM can have its parameters derived immediately from such maps. Exemplary spectra and its position on the maps is shown with green dot. }
\label{fig:4}
\end{figure}

\subsection{Gaia Science Alerts}
Photometric measurements collected by the satellite will be available
for the first analysis after about 24 hours. Science Alerts are
responsible for rapid detection of flux anomalies in the initial data
caused by, e.g. supernovae, dwarf novae or microlensing events.
Science Alerts tools will also analyse the accumulated data and will
use SOMs to detect changes and anomalies in the spectra of the sources
due to, for example, variability in eclipsing binaries or pulsating
stars. Additionally, SOM will immediately allow for detection of new
kind of spectra, not similar to any of known spectra \citep{EvansBelokurovSOM}.

The SOM was trained with about 8000 spectra from the Basel library
(\cite{Basel}) covering wide ranges of temperature, surface gravity
and metallicity. These parameters were then mapped back on the map
(see Figure \ref{fig:4}). Multidimensional sorting ability of the SOM
can be easily seen. With a SOM trained, for any spectrum a winning
node can be found (i.e. the most similar spectrum) and its
temperature, surface gravity and metallicity can be simply read out
from the maps.  Gaia Science Alerts will be tracking changes of the
physical parameters of sources and will alert on any anomalous
behaviour. Also shown an exemplary spectrum with known logTeff=3.9,
logg=3.5 and Fe/H=-0.15, which was identified to be the most similar
to the node (20,3) of the SOM. Its position on parameters' maps is
marked with green dots. It corresponds exactly to the known parameters of the spectrum. 
This useful feature of SOMs can have wide and numerous applications in rapid classification of spectra. 

\section{Applications of SOMs on light curves}

\begin{figure}
  \includegraphics[height=.36\textheight]{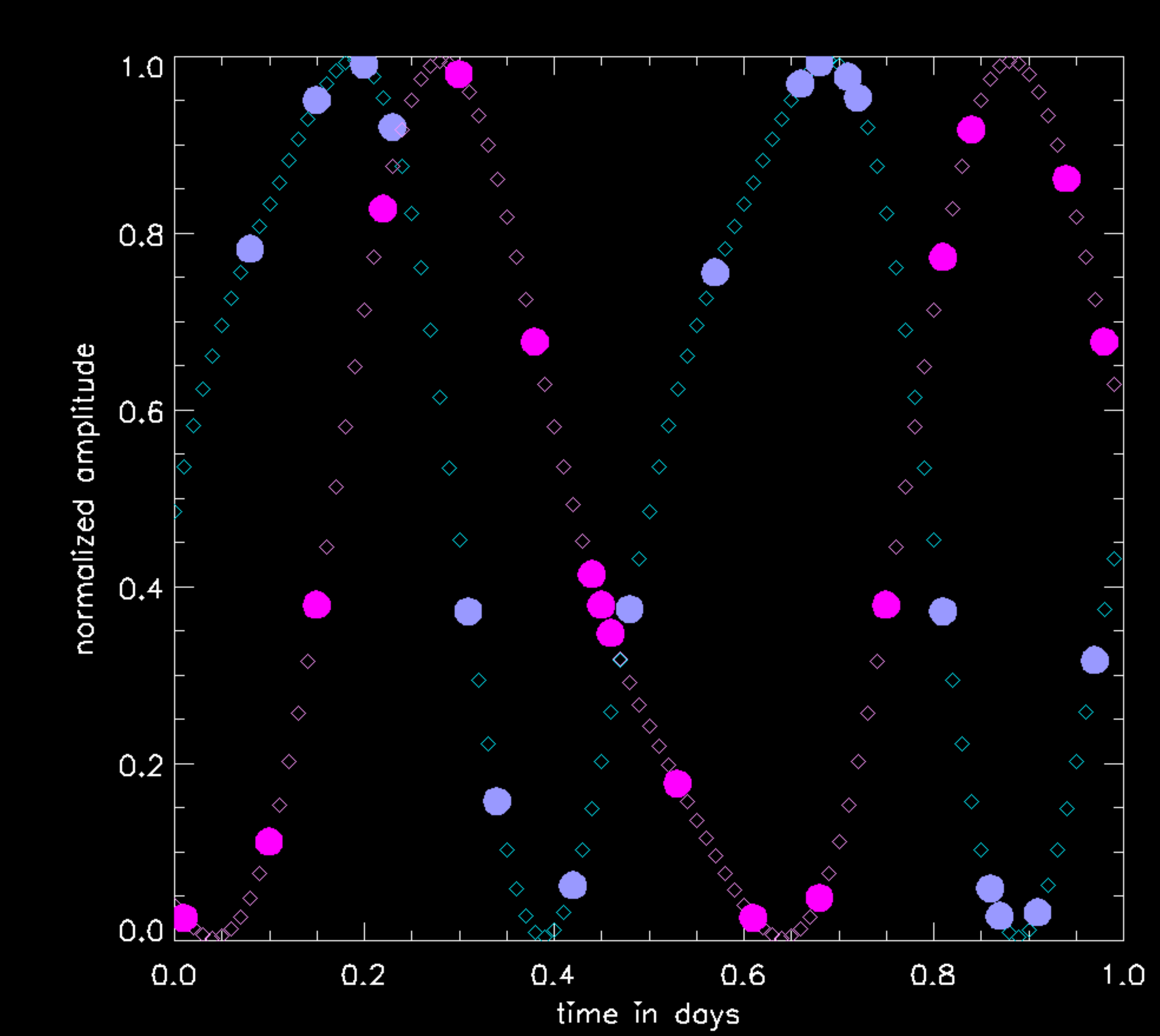}
  \caption{SOM can fill the gaps (empty points) between sparsely sampled data (filled points). }
\label{fig:5}
\end{figure}

Future astronomical surveys will be observing billions of stars. The
light curves of millions of variable stars sharing a similar shape
(e.g. RR Lyrae, eclipsing binaries) will be observed at different
phases. SOM can be used for completing the missing parts of the light
curve when the sampling is not frequent enough but the number of
available light curves is large (like in the case of Gaia's
photometric data stream). This SOM was trained with two different
RR-Lyra-like simulated variables observed sparsely at random phases
(big dots on Figure \ref{fig:5}). After training with several hundreds
of patterns, such SOM is now capable of distinguishing between the two
types of variables and can fill the gaps between the observed data
points.

\begin{figure}
  \includegraphics[height=.36\textheight]{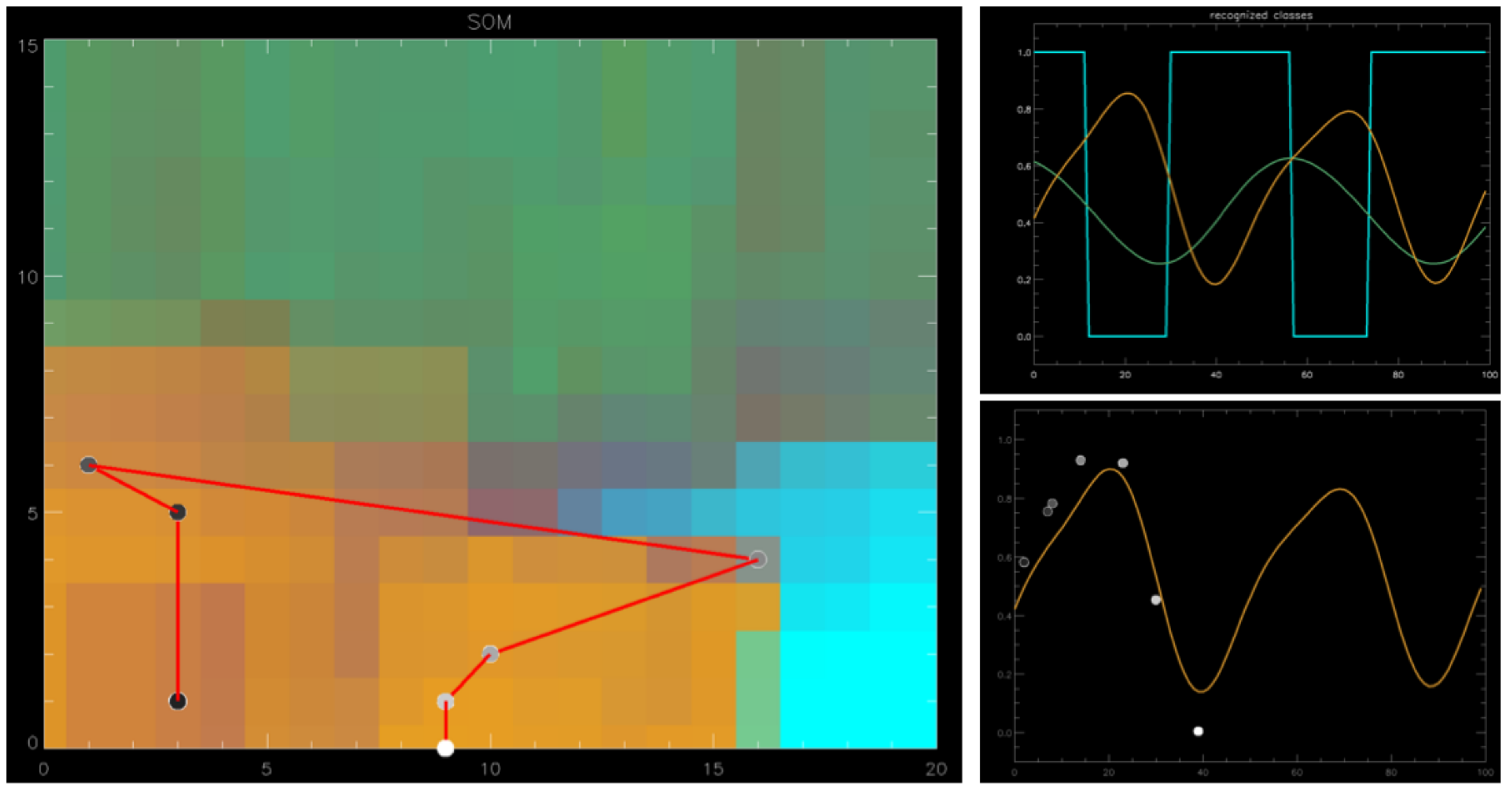}
  \caption{SOM trained on three different type of variability (upper right panel) with a
    track showing convergence of recognition of incomplete data. Black-grey-white dots are consequent data points added to the light curve(lower right panel).}
\label{fig:6}
\end{figure}

Another SOM shown in Figure \ref{fig:6} was trained with patterns
comprising of light curves of three different kinds of simulated
variable stars (upper right panel). As in real world, every light
curve was shifted to start at a random phase. To deal with such
offsets, this SOM first uses a cross-correlation function to match the
phase of the pattern and only then uses the Euclidian distance to
measure similarity. This novelty approach is important in training
SOMs with light curves and has a great potential in classifying
variable stars in real data sets.

The SOM trained to recognise three types of variable light curves
(above) is also capable of figuring out the class of a new data
pattern using only a few first data points. Grey shaded points in
Figure \ref{fig:6} show changes in the classification as more data
points are being added to the input pattern. Each incomplete pattern
was shown to the map and the winner was found. The track (in red on
the map) converged to the correct answer (node 9,0) already when only
7 points of the light curve were present.  This SOM feature can be
applied for real-time classification of variable stars in incoming
data.

\bibliographystyle{aipproc}   
\bibliography{wyrzykowski}

\end{document}